\documentclass{ragtime}

\title[Recurrence analysis of spinning particles in Schwarzschild]
{Recurrence analysis of spinning particles in the Schwarzschild background}

\author[O. Zelenka, G. Lukes-Gerakopoulos and V. Witzany]
       {Ond\v{r}ej Zelenka\at[]{1,2,a},
       Georgios Lukes-Gerakopoulos\at[]{1,b}
        \splitauthors and  Vojt\v{e}ch Witzany\at[]{1,c} \\
        \ins{1}Astronomical Institute of the Academy of Sciences of the Czech Republic,\splitins[1]%
        Bo\v{c}n\'{i} II 1401/1a, CZ-141 31 Prague, Czech Republic\\
        \ins{2} Institute of Theoretical Physics, Faculty of Mathematics and Physics,
        \splitins[2] Charles University, CZ-180 00 Prague, Czech Republic\\
        \ins{a}\Email{ondrzel@gmail.com},
        \ins{b}\Email{gglukes@gmail.com},
        \ins{c}\Email{witzany@asu.cas.cz}
       } 

\begin{document}

\begin{abstract}
   In this work the dynamics of a spinning particle moving in the Schwarzschild background is studied. In particular, the methods of Poincar\'{e} section and recurrence analysis are employed to discern chaos from order. It is shown that the chaotic or regular nature of the orbital motion is reflected on the gravitational waves. 
\end{abstract}


\section{Introduction}

The equations of motion of a small extended test body in curved spacetimes  were first derived by  \citet{Mathisson:1937} and \citet{Papapetrou:1951}, and later reformulated by \citet{Dixon:1970a,Dixon:1970b,Dixon:1974}.  The study of such bodies is usually  reduced to the pole-dipole approximation, in which all the higher-order multipoles are neglected. In this approximation the test body is characterized solely by its mass and spin and it is called a spinning particle. When this particle is subject only to the gravitational interaction, the equations of motion of the particle read
\begin{align}
\frac{{\rm D}P^{\mu}}{{\rm d}\tau} & =-\frac{1}{2}\,{R^{\mu}}_{\nu\kappa\lambda}v^{\nu}S^{\kappa\lambda} \quad,\label{eq:MomEOM}\\
\frac{{\rm D}S^{\alpha\beta}}{{\rm d}\tau} & = P^{\alpha}v^{\beta}-v^{\alpha}P^{\beta}\quad,\label{eq:SpinEOM}
\end{align}
where $P^{\mu}$ denotes the four-momentum, $S^{\mu\nu}$ denotes the spin tensor, $v^{\mu} = \mathrm{d}x^\mu/\mathrm{d}\tau$ denotes the four-velocity (we choose the affine parameter $\tau$ to be the proper time), and ${R^{\mu}}_{\nu\kappa\lambda}$ denotes the Riemann tensor. This set of equations is often called the Mathisson-Papapetrou-Dixon (MPD) equations. To be able to evolve the MPD equations, one has to fix the center of the mass of the body $x^\mu$ by imposing a so called Spin Supplementary Condition (SSC). The SSC, we have implemented in this work, is the Tulczyjew--Dixon (TD) \citep{Tulczyjew:1959,Dixon:1970a} one
\begin{align}\label{eq:TD_SSC}
  S^{\mu\nu} P_{\nu}=0 \quad .
\end{align}
For this SSC the 4-velocity is related to the other variables through:
\begin{align}
v^{\mu}=\frac{m}{\mu^{2}}\left(P^{\mu}+\frac{2S^{\mu\nu}R_{\nu\iota\kappa\lambda}P^{\iota}S^{\kappa\lambda}}{4\mu^{2}+R_{\alpha\beta\gamma\delta}S^{\alpha\beta}S^{\gamma\delta}}\right)  \quad, \label{eq:P_U_TD}
\end{align}
where $\mu^2=-P_\nu P^\nu$ is the mass defined with respect to the momentum and $m=-P_\nu v^\nu$ is the mass defined with respect to the velocity. 

In the case of TD SSC, $\mu$ is a constant of motion independently from the spacetime background. This holds also for the measure of the spin $S=\frac{1}{2}S_{\mu\nu}S^{\mu\nu}$. There are, however, some  background-dependent constants constructed from Killing vectors.
In particular, for a Killing vector $\xi^\mu$ the quantity 
\begin{align} \label{eq:ConsMot}
 C=\xi^\mu P_\mu-\frac12 \xi_{\mu;\nu} S^{\mu\nu}
\end{align}
remains conserved along the worldline $x^\mu(\tau)$ \citep{Dixon:1970a}. In the case of the Schwarzschild spacetime the integrals are four. Namely the energy $E$ and the three components of the total angular momentum $J_b=(J_x,J_y,J_z)$. In the case of geodesic motion, which corresponds to the case $S=0$, the respective system is integrable, since for the respective Hamiltonian $\displaystyle{H= g^{\mu\nu} P_\mu P_\nu/(2 \mu)}$ there are as many degrees of freedom as integrals. In particular, there is the energy, two components of total angular momentum\footnote{The components of the total angular momentum are not mutually in involution, thus from the three components only the two could be taken into account.} and the preservation of the Hamiltonian function itself $H=-\mu/2$. The introduction of the spin increases the degrees of freedom cancelling the integrability and induces chaotic motion to the system \citep{Suzuki:1997}. \citet{Witzany:2019} showed that, independent of the space-time background, the spinning particle under the TD SSC has only one additional active degree of freedom as compared to the geodesic problem (the structureless test particle), at least if the conservation of the spin measure as well as the TD constraint itself are taken into account. This implies that by using the remaining constants of motion ($E$ and two components of $J_b$) in the case of the Schwarzschild background, the degrees of freedom can be reduced to two, i.e., the dynamics of the system can be described in a 4 dimensional phase space. 

This work revisits the study of chaos in the case of a spinning particle moving in the Schwarzschild spacetime, which was for the first time performed by \citet{Suzuki:1997}. Since the dynamics of the studied system can be confined to 2 degrees of freedom by fixing the values of the integrals of motion, a 2D Poincar\'{e} section is an accurate method to study the dynamics of the system. However, when the number of degrees of freedom is higher than 2, such as for a spinning particle moving in a Kerr background, then a 2D Poincar\'{e} section is not a reliable method to study the dynamics \citep[see, e.g.,][]{LGKPS:2016}. For studying systems independently from the number of degrees of freedom {\it recurrence analysis} is considered to be a more appropriate method \citep[see, e.g., the review of][~and reference therein]{Marwan:2007}. Further advantage of recurrence analysis is that it is a method analyzing time series, which is advantageous when we consider signals from gravitational wave strains later on. Thus, in this work we test the performance of the  recurrence analysis by comparing it with the standard method of a 2D Poincar\'{e} section.

\paragraph*{Units and notation:} Geometric units are used throughout the article, ${G=c=1}$. Greek letters denote the indices corresponding to spacetime, while Latin letters denote indices corresponding only to space. We use the Riemann tensor defined as
${{R^\alpha}_{\beta\gamma\delta}=
 \Gamma^\alpha_{\gamma \lambda} \Gamma^\lambda_{\delta \beta}
 - \partial_\delta \Gamma^\alpha_{\gamma\beta}
 - \Gamma^\alpha_{\delta\lambda} \Gamma^{\lambda}_{\gamma\beta}
 + \partial_\gamma \Gamma^{\alpha}_{\delta \beta}}$,
where the Christoffel symbols $\Gamma$ are computed
from the metric with signature $(-,+,+,+)$. The Levi-Civita tensor is $\epsilon_{\mu\nu\rho\sigma}=\sqrt{-g}\tilde{\epsilon}_{\mu\nu\rho\sigma}$, with the Levi-Civita symbol $\tilde{\epsilon}_{0123}=1$.

\section{Comparing Poincar\'{e} section method with recurrence analysis}

 \begin{figure}[t]
 \begin{center}
 \includegraphics[width=0.7\linewidth]{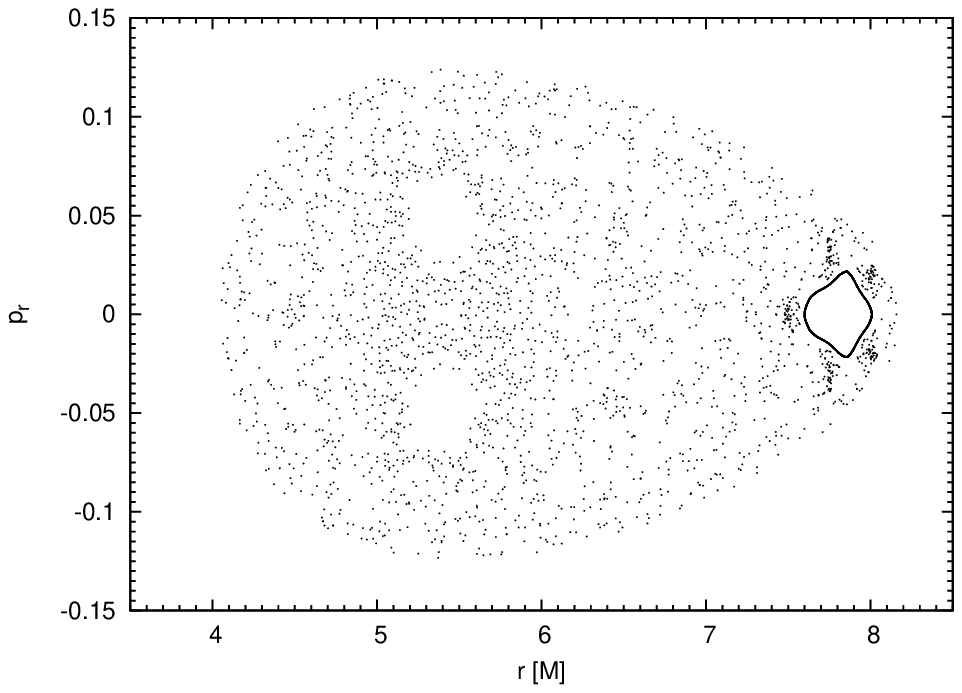}
 \includegraphics[width=0.45\linewidth]{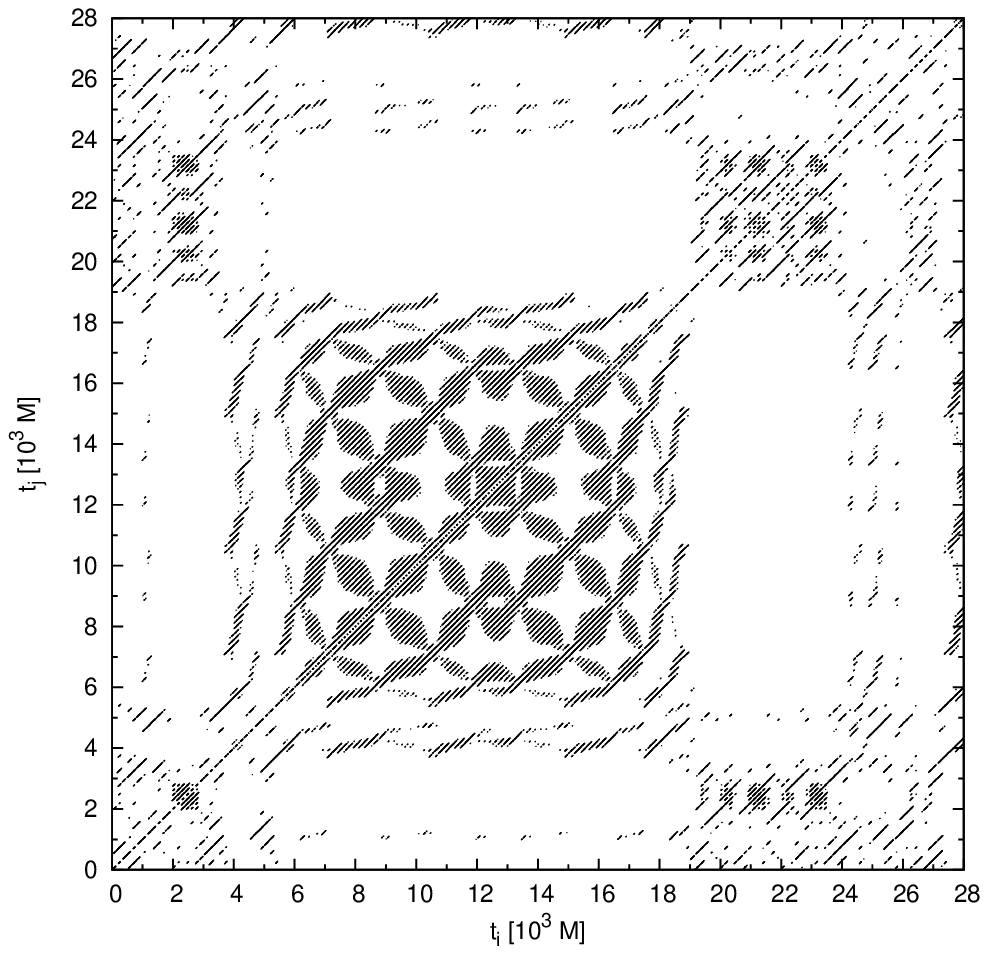}
 \includegraphics[width=0.45\linewidth]{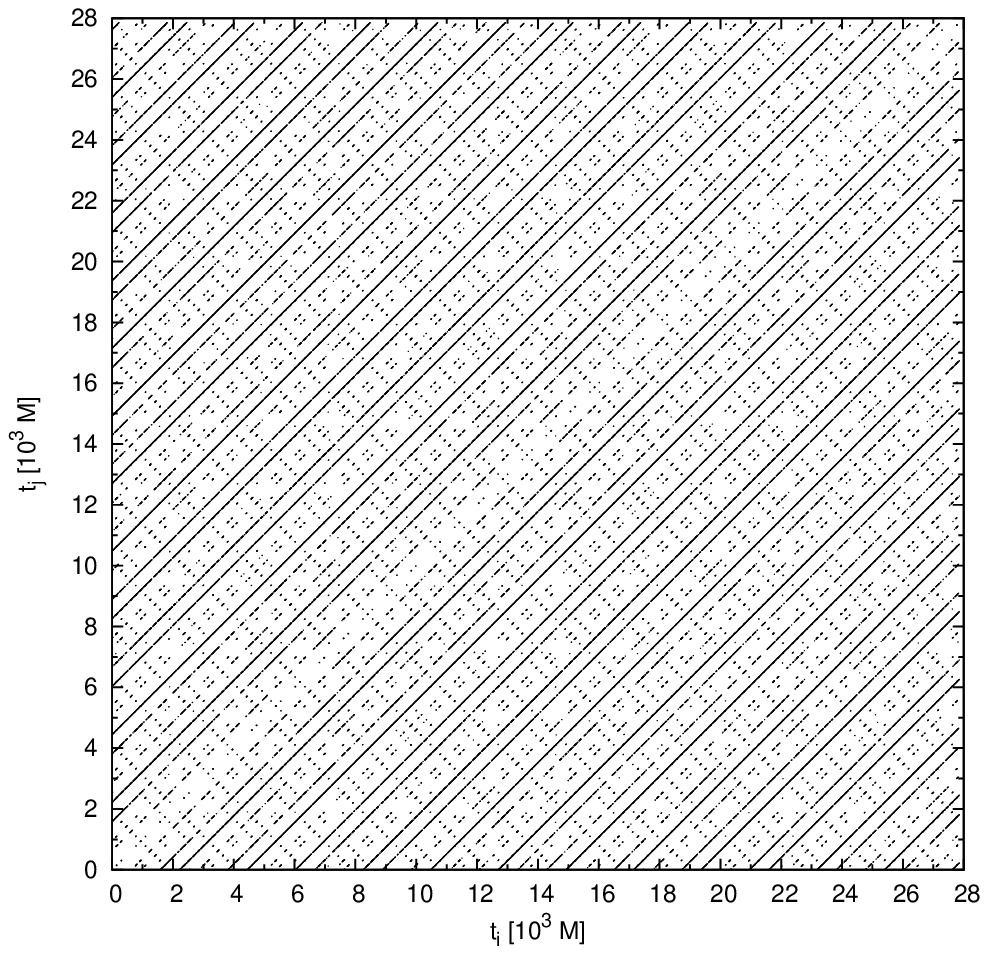}
  \end{center}
 \caption{\textit{Top panel}: A Poincar\'e section on the equatorial plane $\theta=\pi/2$ with $P_\theta>0$, $E=0.92292941\mu$, $J_z=4.0\mu M$, $S=1.4\mu M$. \textit{Bottom left panel}: The recurrence plot for a chaotic trajectory with initial conditions $r=4.5M$, $P_r=0$; recurrence threshold $\varepsilon=0.87083$. \textit{Bottom right panel}: The recurrence plot for a regular trajectory with initial conditions $r=7.6M$, $P_r=0$; recurrence threshold $\varepsilon=0.49013$.}
 \label{fig:PoiVsRA} 
 \end{figure}

According to the recurrence analysis, if $\boldsymbol{y}(t)$ is a vector time series in an arbitrary phase space, then a \textit{recurrence} occurs when the distance between the $i$th point and the $j$th point of the time series drops below a  threshold $\varepsilon$. These recurrences are recorded in the recurrence matrix 
 \begin{align} \label{eq:RecMat}
   \mathcal{R}(i,j;\varepsilon)=\Theta(\varepsilon-||\boldsymbol{y}(i)-\boldsymbol{y}(j)||)\quad,
 \end{align}
 where $||.||$ denotes a norm in the phase space and $\Theta$ denotes the Heaviside step-function. A depiction of a recurrence matrix produces a recurrence plot \citep[see, e.g.,][]{Marwan:2007}. By inspecting a recurrence plot, as by inspecting a Poincar\'{e} section, one can tell whether a time series is chaotic or not. On a Poincar\'{e} section a chaotic orbit appears as a swarm of scattered points, an example of which can be seen in the top panel of Fig.~\ref{fig:PoiVsRA}. On the other hand, on a recurrence plot a chaotic orbit can be identified by observing square--like structures, as can be seen in the left bottom panel of Fig.~\ref{fig:PoiVsRA}. A regular orbit is depicted on a Poincar\'{e} section as a smooth zero-width closed curve, as the one lying at $7\lesssim r \lesssim 8 $ in the top panel of Fig.~\ref{fig:PoiVsRA}, while on a recurrence plot the regularity of the orbit manifests itself by long diagonal lines covering the whole plot.
 
 For the initial conditions of Fig.~\ref{fig:PoiVsRA} we have followed the setup suggested by \cite{Suzuki:1997}. Namely, we have chosen $J_z$ to be the only non-zero total angular momentum component, i.e. $J_b=(0,0,J_z)$; we have fixed the energy $E$ and the spin measure $S$, which is most conveniently expressed in units of $\mu M$, where $M$ is the mass of the central Schwarzschild black hole. Apart from the constants, we always choose initial conditions such that $\theta=\pi/2, P_r=0$ and $r$ varying from orbit to orbit ($t,\phi,r,\theta$ are the usual Schwarzschild coordinates). From the four components of the SSC (Eq.~\eqref{eq:TD_SSC}) only three are linearly independent, and along with the choice of the constants of motion this setup determines the six components of the spin tensor and the remaining three components of the momentum. For more details on how to set up the initial conditions the interested reader is referred to \cite{Suzuki:1997}. 
 
 To evolve the MPD equations with TD SSC one has to use Eq.~\eqref{eq:P_U_TD} at each integration step and take into account the fact that $v^\mu v_\mu=-1$. This procedure actually fixes the mass $m$ at each integration step. The time series for the recurrence plots in Fig.~\ref{fig:PoiVsRA} were obtained by the method explained in Appendix~\ref{sec:app_rp}.
 
 \section{Gravitational wave strains}  

 \begin{figure}[t]
 \begin{center}
 \includegraphics[width=0.7\linewidth]{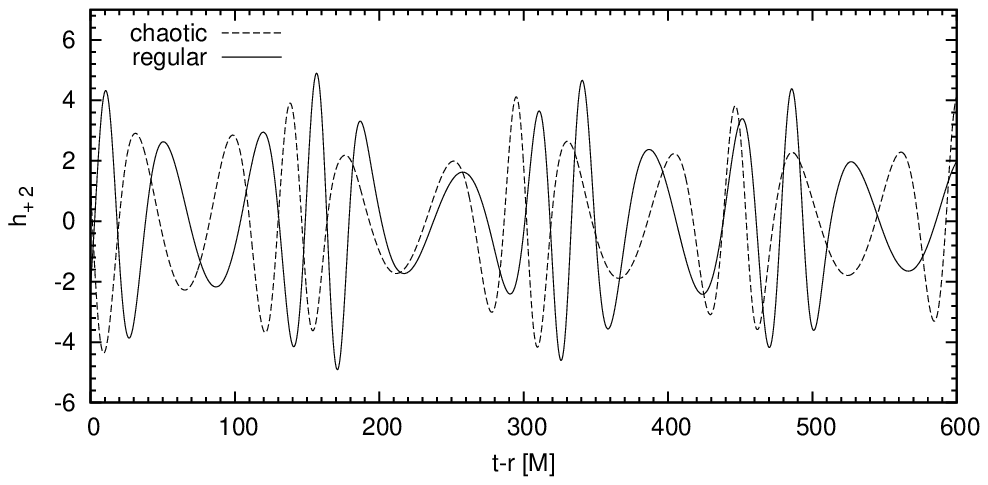}
 \includegraphics[width=0.45\linewidth]{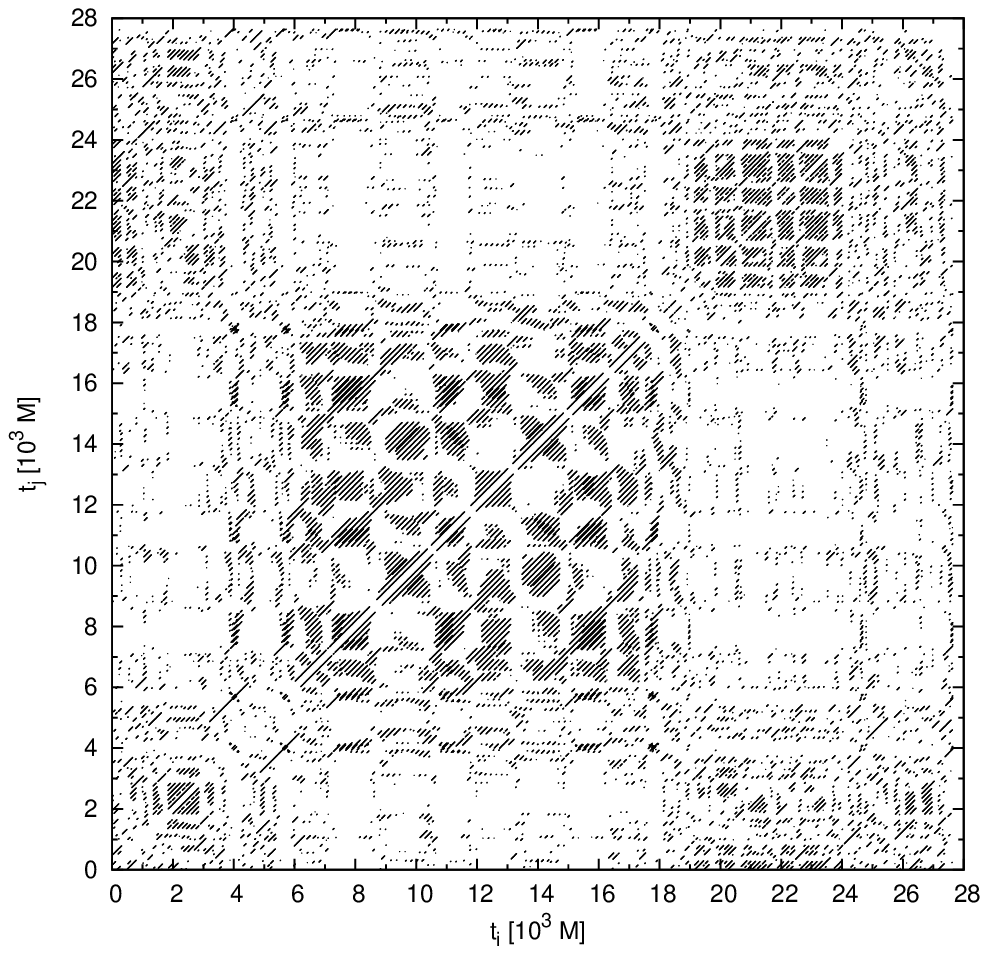}
  \includegraphics[width=0.45\linewidth]{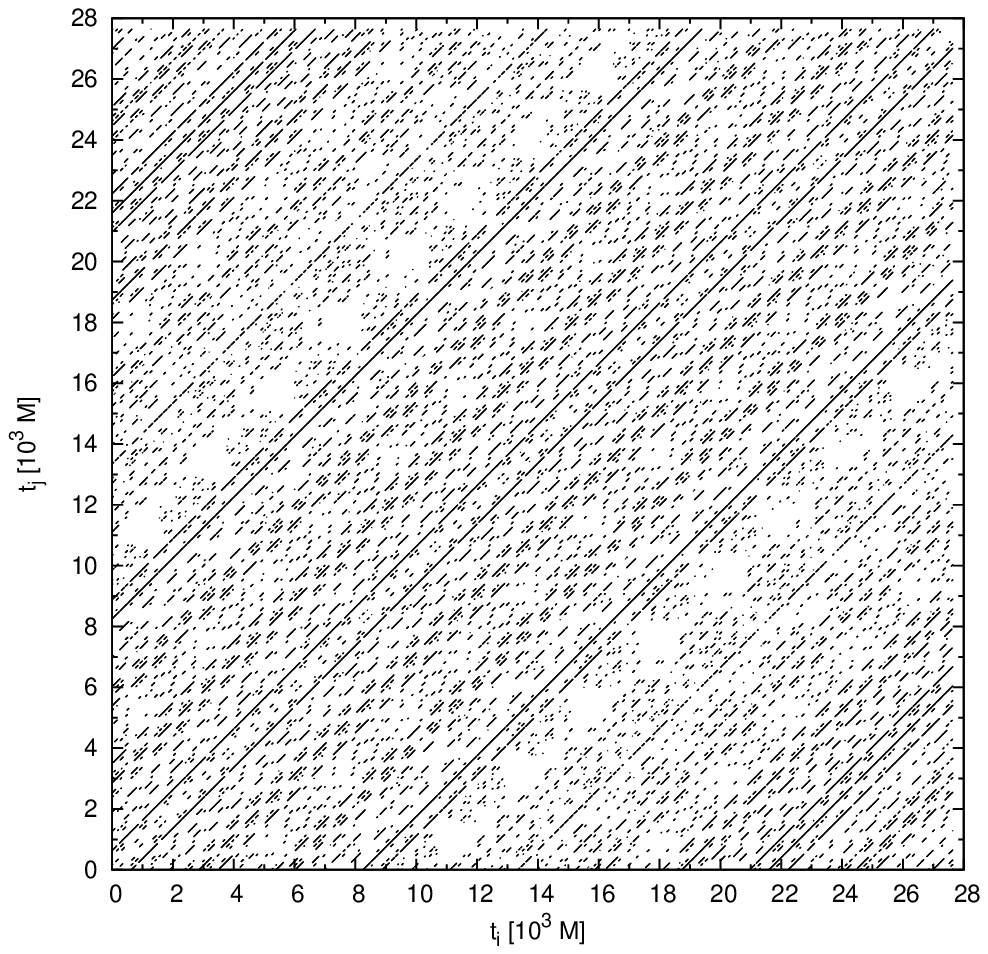}
  \end{center}
 \caption{\textit{Top panel}: The gravitational waveforms of the strain mode ${h_{+}}_2$  corresponding to the orbits presented in Fig.~\ref{fig:PoiVsRA}.  \textit{Bottom left panel}: The recurrence plot of the waveform corresponding to the chaotic orbit, using time delay $8.664M$ and embedding dimension 21, $\varepsilon=8.566$. \textit{Bottom right panel}: The recurrence plot for the waveform corresponding to the regular orbit, using time delay $8.664M$ and embedding dimension 21, $\varepsilon=6.819$.
 }\label{fig:GWRA} 
 \end{figure}

 In this section we will discuss whether chaos and order can be discerned in gravitational waves. We shall use gravitational waves emitted from a spinning particle moving in the Schwarzschild background. In a similar study, \citet{Kiuchi:2004} have used the analytic formula of multipole expansion of gravitational field to calculate the gravitational waves. In our study, we use a time-domain Teukolsky equation solver called Teukode. Teukode was developed by \cite{Harms:2014} and in \citet{Harms:2016} the spin of the particle was incorporated.
 
 From Teukode we obtain the strain $h_+$ decomposed in a spin-weighted spherical harmonic basis
 \begin{align}
     h_{+}=\sum_{\text{m}=1}^\infty {h_+}_\text{m} =
     \sum_{l=2}^{\infty}\sum_{\text{m}=1}^{\text{m}=l} {h_{+}}_{l\text{m}}.
 \end{align}
 For the purposes of our study we use just ${h_+}_2$. The waveforms of the strain for the two cases in the bottom panels of Fig.~\ref{fig:PoiVsRA} are shown in the top panel of Fig.~\ref{fig:GWRA}. From looking at the shapes of the waveforms alone one cannot tell whether they belong to a chaotic or a regular trajectory, which is in agreement with the findings of \citet{Kiuchi:2004}. To get an answer to the above issue one has to apply an appropriate chaos detection technique. In our work this technique is the recurrence analysis. In the bottom panels of Fig.~\ref{fig:GWRA}, we see recurrence plots of ${h_+}_2$, the left corresponds to gravitational waves from the chaotic orbit and the right corresponds to gravitational waves from the regular orbit of Fig.~\ref{fig:PoiVsRA}. 
 
 The recurrence plots of  Fig.~\ref{fig:GWRA} look quite similar to the respective ones in Fig.~\ref{fig:PoiVsRA}, thus they characterize the orbits in the same way as in Fig.~\ref{fig:PoiVsRA}. Namely, the left bottom panel is dominated by square-like structures indicating chaos and the right bottom panel is dominated by diagonal lines indicating order. In conclusion, the information about the chaoticity or the regularity of an orbit is encoded in the respective gravitational waves. 
 
 In the regular case of the right panel of Fig.~\ref{fig:GWRA} a more careful inspection shows that the diagonal lines are slightly diffused. This diffusion is introduced by the numerical accuracy of Teukode. This is similar to what happened when \citet{LGK:2018} polluted the time series with white noise. Moreover, it should be mentioned that this is the first time that Teukode has been tested for off-equatorial orbits. The fact that the orbital and the waveform recurrence plots do not only indicate the same dynamical nature, but actually look alike, confirms that Teukode is performing well also for off-equatorial orbits.

\section{Summary}

 We have employed recurrence analysis to discern chaos from order in the case of a spinning particle moving in the Schwarzschild background. In particular, we have first provided a Poincar\'{e} section, on which we identified one regular and one chaotic orbit. For these two orbits we have produced the respective recurrence plots and we have confirmed their nature with respect to chaoticity. Then, we fed these two trajectories to the Teukode to produce the respective gravitational waveforms. Since from just inspecting a waveform one cannot tell whether it comes from a regular or chaotic trajectory \citep{Kiuchi:2004}, we have applied recurrence analysis on the gravitational waveforms. The waveform recurrence plots and the respective orbital ones look very similar, which indicates that the information about the chaoticity or not of an orbit can be revealed in the emitted gravitational waves.

\ack

The authors are supported by Grant No. GA\v{C}R-17-06962Y of the Czech Science Foundation. G.L.-G. would like to acknowledge networking support by the COST Action CA16104. O.Z. and G.L.-G. would also like to express gratitude for the hospitality of the Theoretical Physics Institute at the University of Jena. Finally, we would like to thank Ond\v{r}ej Kop\'{a}\v{c}ek, Sebastiano Bernuzzi, Enno Harms and Sarp Akcay for useful discussions and comments. 
\bibliography{RAGtime20_RA}

\appendix
\section{Recurrence plots}\label{sec:app_rp}
The recurrence plots for the trajectories in Fig.~\ref{fig:PoiVsRA} have been produced using the following method: points of the numerically integrated trajectory were sampled at a rate of $\Delta t = 8.664M$ and the data for $r$, $P_r$, $\theta$, $P_\theta$, $S^t$, $S^r$, $S^\theta$, $S^\phi$ $(S^\mu \equiv -\frac{1}{2} \epsilon^{\mu\nu\rho\sigma} \, P_\nu \, S_{\rho\sigma}/\mu)$ were extracted. Each of these 8 time series was rescaled to have zero mean and unit variance. This way, we obtained data in an 8-dimensional space and computed the recurrence matrix using Eq.~\eqref{eq:RecMat} with the Euclidean metric.

Computation of the recurrence plots of gravitational waveforms in Fig.~\ref{fig:GWRA} was slightly more complicated, because in this case there is only limited information available (we used the strain $h_{+~2}$) as opposed to full phase space vectors when working with trajectories. It is therefore necessary to use some technique of phase space reconstruction, in this case the time delay method. We provide a short description of the method; for more details, the reader is referred to \citet{Marwan:2007}.

The time delay method has been proven to provide a diffeomorphism between the original and the reconstructed phase space under certain assumptions. Consider a time series $\boldsymbol{x}(t)$. The reconstructed time series vector is then
\begin{equation}
    \boldsymbol{y}(t) = \left(\boldsymbol{x}(t), \boldsymbol{x}(t+\Delta t),\hdots,\boldsymbol{x}(t+(N-1)\Delta t)\right)\quad,
\end{equation}
where $\Delta t$ is called the time delay and $N$ is the embedding dimension. Both of these are essentially free parameters, but there are methods to fix these for optimal results. The canonical choice of the time delay is the first minimum of the mutual information. To obtain a reasonable embedding dimension one can study the fraction of false nearest neighbors, that is, the fraction of points whose nearest neighbor in the reconstructed phase for the given embedding dimension becomes more distant by a certain factor when the dimension is increased.

\end{document}